\documentclass[useAMS,usenatbib]{mn2e}

\usepackage{epsfig,graphicx,mathptm}
\usepackage{times}
\usepackage{natbib}
\usepackage{amsmath}
\DeclareMathOperator\erf{erf}
\DeclareMathOperator\Log{Log}

\newcommand{\gtrsim}{\ga}
\newcommand{\lesssim}{\la} 
\newcommand{\mone}{$^{-1}$}
\newcommand{\sqr}{$^2$}
\newcommand{\cub}{$^3$}
\newcommand{\flunits}{erg s$^{-1}$ cm$^{-2}$}
\newcommand{\ypar}{$y$-parameter}
\newcommand{\lcdm}{$\Lambda$CDM}
\newcommand{\lcdmn}{\lcdm$\nu$}
\newcommand{\omegam}{$\Omega_{\rm m}$}
\newcommand{\omegacdm}{$\Omega_{\rm CDM}$}
\newcommand{\omegal}{$\Omega_\Lambda$}

\newcommand{\omeganu}{$\Omega_\nu$}
\newcommand{\fnu}{$f_\nu$}
\newcommand{\smnu}{$\Sigma \, m_\nu$}
\newcommand{\smnut}{$\Sigma \, m_\nu=(0,0.17,0.34)$ eV}
\newcommand{\sigmae}{$\sigma_8$}
\newcommand{\hzero}{$H_0$}
\newcommand{\gadget}{\textsc{gadget}}
\newcommand{\gadgetthree}{\textsc{gadget-3}}
\newcommand{\subfind}{\textsc{subfind}}
\newcommand{\msun}{$M_{\sun}$}
\newcommand{\mfiveh}{$M_{500}$}
\newcommand{\rfiveh}{$R_{500}$}
\newcommand{\yfiveh}{$Y_{500}$}
\newcommand{\hmone}{$\,h^{-1}$}
\newcommand{\hmthree}{$\,h^{-3}$}
\newcommand{\wmap}{{\it WMAP}}
\newcommand{\planck}{{\it Planck}}
\newcommand{\erosita}{{\it eROSITA}}
\newcommand{\spt}{SPT}
\newcommand{\act}{ACT}

\def\prd{Phys. Rev. D}
\def\aap{A\&A}
\def\apj{ApJ}

\def\mnras{MNRAS}

\def\physrep{Phys. Rep.}

\def\apss{Ap\&SS}      % Astrophysics and Space Science
\def\apjs{ApJS}
\def\jcap{JCAP}        % Journal of Cosmology and Astro-Particle Physics

\title[Massive neutrinos in SZ and X-ray]
{The effect of massive neutrinos on the Sunyaev--Zel'dovich and X-ray observables of 
galaxy clusters}
\author[M. Roncarelli et al.]
{M. Roncarelli$^{1,2}$\thanks{E-mail: mauro.roncarelli@unibo.it},
C. Carbone$^{3,4}$ and  
L. Moscardini$^{1,2,4}$
\\
$^1$Dipartimento di Fisica e Astronomia, Universit\`a di Bologna, viale Berti Pichat 
6/2, I-40127 Bologna, Italy \\
$^2$Istituto Nazionale di Astrofisica (INAF) -- Osservatorio Astronomico di Bologna, via 
Ranzani 1, I-40127 Bologna, Italy \\
$^3$Istituto Nazionale di Astrofisica (INAF) -- Osservatorio Astronomico di Brera, via 
Bianchi 46, I-23807 Merate (LC), Italy \\
$^4$Istituto Nazionale di Fisica Nucleare (INFN) -- Sezione di Bologna, 
viale Berti Pichat 6/2, I-40127 Bologna, Italy
}

\begin{document}

\date{Accepted 2014 December 1. Received 2014 November 5; in original form 2014 September 
12}

\pagerange{\pageref{firstpage}--\pageref{lastpage}} \pubyear{2014}

\maketitle

\label{firstpage}

\begin{abstract}
Massive neutrinos are expected to influence the formation of the large-scale structure of 
the Universe, depending on the value of their total mass, \smnu. In particular \planck\ 
data indicate that a non-zero \smnu\ may help to reconcile CMB data with 
Sunyaev-Zel'dovich (SZ) cluster surveys. In order to study the impact of neutrinos on the 
SZ and X-ray cluster properties we run a set of six very large cosmological simulations 
(8\hmthree\ Gpc\cub\ comoving volume) that include a massive neutrino particle component: 
we consider the values of \smnut\ in two cosmological scenarios to test possible 
degeneracies. Using the halo catalogues extracted from their outputs we produce 50 mock 
light-cones and, assuming suitable scaling relations, we determine how massive neutrinos 
affect SZ and X-ray cluster counts, the \ypar\ and its power spectrum. We provide 
forecasts for the South Pole Telescope (\spt) and \erosita\ cluster surveys, showing that 
the number of expected detections is reduced by 40 per cent when assuming \smnu=0.34 eV 
with respect to a model with massless neutrinos. However the degeneracy with \sigmae\ and 
\omegam\ is strong, in particular for X-ray data, requiring the use of additional probes 
to break it. The \ypar\ properties are also highly influenced by the neutrino mass 
fraction, \fnu, with $\langle\,y\,\rangle\propto(1-f_\nu)^{20}$, considering the cluster 
component only, and the normalization of the SZ power spectrum is proportional to 
$(1-f_\nu)^{25-30}$. Comparing our findings with \spt\ and Atacama Cosmology Telescope 
measurements at $\ell=3000$ indicates that, when \planck\ cosmological parameters are 
assumed, a value of \smnu$\simeq0.34$ eV is required to fit with the data.
\end{abstract}


%%%%%%%%%%%%%%%%%%%%%%%%%%%%%%%%%%%%%%%%%%%%%%%%%%%%%%%%%%%%%%%%%%%%%%%%%%%%%%%%%%%%%%%%%%
%%%%%%%%%%%%%%%%%%%%%%%%%%%%%%%%%%%%%% Introduction %%%%%%%%%%%%%%%%%%%%%%%%%%%%%%%%%%%%%%
%%%%%%%%%%%%%%%%%%%%%%%%%%%%%%%%%%%%%%%%%%%%%%%%%%%%%%%%%%%%%%%%%%%%%%%%%%%%%%%%%%%%%%%%%%

\section{Introduction} \label{sec:intro}
The Standard Model (SM) of particle physics predicts the existence of three active 
neutrino species: the electron ($\nu_{\rm e}$), muon ($\nu_\mu$) and tau ($\nu_\tau$) 
neutrinos. These leptonic particles are chargeless and interact only via the weak force 
making them very elusive and difficult to study and leaving many open questions about 
their physical properties and on the possible existence of additional sterile species 
\citep[see][for a review]{mohapatra07}. While the SM originally assumed neutrinos to be 
massless, the discovery of leptonic flavour oscillations suggests that they have a 
non-zero mass, fixing the lower limit for the sum of neutrino masses to 
\smnu$\equiv m_{\nu_e}+m_{\nu_\mu}+m_{\nu_\tau}\gtrsim 0.05$ eV\footnote{
More specifically, \smnu\ must be greater than approximately $0.06$ eV in the 
normal hierarchy scenario and $0.1$ eV in the degenerate hierarchy.
}
\citep[see][and references therein]{lesgourgues06,lesgourgues12,lesgourgues14,
lesgourgues13,gonzalezgarcia14}. This opens the possibility to study neutrino properties 
also via their gravitational interaction.

From the cosmological point of view the presence of a thermal neutrino component has two 
important effects. First, since they become non-relativistic after the epoch of 
recombination they behave as an additional component in the radiation-dominated era, 
modifying the radiation density term
\begin{equation}
\rho_r = \rho_\gamma+\rho_\nu=
\left[1+\frac{7}{8}\left(\frac{4}{11}\right)^\frac{4}{3}N_{\rm eff}\right]\rho_\gamma\, ,
\end{equation}
where $\rho_\gamma$ and $\rho_\nu$ are the photon and neutrino energy densities, 
respectively, and $N_{\rm eff}$ is the effective number of neutrino species that 
according to the SM predictions should be fixed to $N_{\rm eff}=3.046$. This has the 
effect of postponing the matter radiation equality for a given value of \omegam$\,h^2$ 
(being \omegam\ the ratio between the matter density of the Universe and the critical 
one, $\rho_{\rm c}$, at $z=0$ and $h$ the Hubble constant $H_0$ in units of 100 km 
s\mone Mpc\mone) and modifying the background evolution, therefore slightly affecting 
the properties of the primary cosmic microwave background (CMB) anisotropies. In 
addition, after recombination neutrinos act as a hot dark matter (HDM) component whose 
energy-density parameter depends on the CMB temperature, $T_{\rm CMB}$, and \smnu\ only:
\begin{equation}
\Omega_\nu \equiv \frac{\rho_\nu}{\rho_{\rm c}} = 
\frac{16\,\zeta(3)\,T_{\rm CMB}^3}{11\upi\,H_0^2}\,\Sigma \, m_\nu \simeq 
\frac{\Sigma \, m_\nu}{93.14\,h^2\,\rm eV} \, ,
\label{e:omnu}
\end{equation}
where\footnote{Equation~(\ref{e:omnu}) is expressed in Planck units, i.e. $G=\hbar=c=
k_{\rm B}=1$.} $\zeta$ is the Riemann zeta function, with $\zeta(3)\simeq1.202$. Even if 
present CMB observations put strong constraints on the amount of HDM in the Universe, 
indicating that \omeganu\ must be very small, this component can produce significant 
effects on the large-scale structure (LSS) evolution. In fact, the large thermal 
velocities of non-relativistic neutrinos suppress the growth of neutrino densities 
perturbations on scales smaller than their characteristic free-streaming comoving length
\begin{equation}
\lambda_{\rm fs,\nu} \simeq 
7.67\,\frac{H_0(1+z)^2}{H(z)}\left(\frac{1\,\rm eV}{m_\nu}\right) h^{-1}\,\rm Mpc.
\label{e:l_fs}
\end{equation}
Since neutrinos induce a gravitational backreaction effect, also the evolution of both 
cold dark matter (CDM) and baryon \citep{rossi14} density is affected and, therefore, the 
total matter power spectrum is hugely suppressed on such scales: this translates into a 
modification of the halo mass function \citep[see e.g.][]{costanzi13b,castorina14,
villaescusa14}. The dependence of equation~(\ref{e:l_fs}) on the neutrino mass makes the 
growth of the LSS a sensitive tool to determine the value of \smnu\ \citep[see e.g.][]
{viel10,marulli11,shimon12,carbone13,costanzi13a,mak13,takeuchi14}.

Indeed, nowadays the tightest upper limits on the total neutrino mass come from 
cosmological studies. In detail, bounds have been put with Lyman-$\alpha$ forest 
observations \citep[e.g.,][]{croft99,viel10}, galaxy redshift surveys \citep{elgaroy02,
tegmark06,thomas10}, CMB observations from the \emph{Wilkinson Microwave Anisotropy 
Probe} \citep[\wmap;][]{komatsu09,komatsu11,hinshaw13}, growth of galaxy clusters 
\citep{mantz10b,mantz15} and galaxy clustering \citep{zhao13,beutler14,sanchez14}. 
Currently, the most stringent upper limit comes from \cite{riemersorensen14} who combine 
the large-scale power spectrum from the WiggleZ Dark Energy Survey 
\citep{riemersorensen12} with CMB observations and measurements of the Baryon acoustic 
oscillations (BAO) scale, yielding the upper limit of \smnu$<0.18$ eV (95 per cent CL).

More recently the attention of the astrophysical community on this problem was raised by 
the \planck\ satellite observations. The cosmological results from the CMB anisotropies 
by \cite{planck14cp} appear to be somewhat in tension with the ones from the thermal 
Sunyaev-Zel'dovich \citep[SZ;][]{sunyaev70} effect galaxy cluster survey obtained with 
the same instruments \citep{planck14cc,planck14sz}: while \planck\ CMB data (combined 
with \wmap\ polarization data and other high-resolution CMB experiments) constrain 
$\sigma_8 (\Omega_{\rm m}/0.27)^{0.3} = 0.87 \pm 0.02$, being \sigmae\ the rms of matter 
density fluctuations in a sphere of 8\hmone Mpc at the present epoch, the galaxy cluster 
counts analysis indicates a significantly lower value of $0.78\pm0.01$, which translates 
into about a factor of 2 less objects detected than expected from the CMB analyses. If on 
one side the systematics connected with both the X-ray and SZ scaling-laws adopted 
\citep[see e.g.][]{sereno14} and the uncertainties in the modelling of non-thermal 
pressure bias \citep[see e.g. the study on \planck\ clusters by][]{vonderlinden14} may 
hamper the robustness of the SZ cluster findings, on the other hand these results are 
consistent with other measurements obtained from other galaxy cluster surveys at 
different wavelengths and affected by completely different systematics 
\citep{vikhlinin09,rozo10,hasselfield13,reichardt13}. Moreover, the low \sigmae\ scenario 
fits also with the expected number counts derived from the \planck\ $y$-parameter map 
analysis \citep{planck14ym}. The presence of massive neutrinos offers a possible natural 
explanation of this apparent discrepancy between the high- and low-redshift universe, 
while CMB analyses assume a six free-parameter standard $\Lambda$ cold dark matter 
(\lcdm) cosmology, adding a small \smnu\ as an additional component can help reconcile 
the discrepancy \citep[see][for a more detailed discussion on this topic]{battye14,
costanzi14}. In the framework of this new \lcdmn\ scenario \cite{planck14cc} obtained 
\smnu$=(0.20\pm0.09)$ eV with a joint analysis of \planck-CMB, \planck-galaxy clusters 
and BAO: if confirmed this result would represent the tightest constraint for neutrino 
masses up to date.

One of the main difficulties in using LSS probes to determine \smnu\ is the degeneracy 
with the other cosmological parameters sensitive to the growth of cosmic structures, 
namely \sigmae\ and \omegam. The key to break this degeneracy is either the 
combination of different observables that allow to determine its redshift evolution, or 
adding CMB priors to the analysis of LSS data. \cite{carbone11} showed that 
future spectroscopic galaxy surveys planned for the next decade, like \emph{Euclid} 
\citep{laureijs11} and \emph{Wide-Field Infrared Survey Telescope} \citep[\emph{WFIRST};]
[]{spergel13}, when combined with \planck\ priors will be able to measure both 
$N_{\rm eff}$ and \smnu\ independently of the galaxy power spectrum normalization, the 
dark energy energy parametrization and the assumption of flat geometry if \smnu$>0.1$ 
eV. Conversely, if \smnu\ is lower, they will allow a measurement in the framework of a 
flat \lcdm\ model. In the next years also galaxy clusters surveys will play an important 
role in this framework \citep{carbone12,costanzi13a}. The South Pole Telescope (\spt) 
team, which has recently released a first catalogue of 224 galaxy clusters detected via 
SZ effect in an area of 720 deg\sqr\ \citep{reichardt13}, is expected to achieve almost 
1000 identifications with the full 2500 deg\sqr\ survey, together 
with a measurement of the SZ power spectrum at the arcminute scale with unprecedented 
accuracy. In addition the \emph{extended Roentgen Survey with an Imaging Telescope Array} 
(\erosita) satellite \citep[see][]{predehl07,merloni12} will perform a full-sky survey in 
the X-rays with the potential of detecting about 10$^5$ galaxy clusters down to $z 
\approx 1-1.5$. Provided that these objects will have adequate redshift measurements, 
these data will improve significantly our knowledge of how the LSS evolves with time. 
This will allow not only an independent set of measurements of both \omegam\ and \sigmae, 
but has also the potential of providing useful constraints on \smnu.

In this work we study how massive neutrinos affect the SZ and X-ray observables of galaxy 
clusters and how the expected detections from upcoming galaxy cluster surveys will be 
influenced by both the value of \smnu\ and by the uncertainty in other cosmological 
parameters. Like in our previous work on non-Gaussianities of primordial density 
fluctuations \citep{roncarelli10a}, we use a set of six cosmological simulations that 
describe the formation of CDM structures including a massive neutrino component, with 
different values of \smnu\ and in two different cosmological scenarios. Starting from 
high redshift, our simulations follow the evolution of a very large comoving volume of 
the Universe, namely 8\hmthree\ Gpc\cub, allowing us to achieve an accurate description 
of the mass function up to the highest mass haloes, well above 10$^{15}$\msun. After 
identifying galaxy clusters in our simulation outputs, we use scaling relations derived 
from observations to link their masses with their SZ and X-ray observables. By 
considering the geometry derived from the corresponding cosmological model, we associate 
to each halo a position in the sky, reconstruct the past light-cones of the LSS and 
create mock \ypar\ observational maps. We compare our results with \planck\ observations, 
both for cluster counts and for the \ypar\ power spectrum, and provide forecasts for 
future \spt\ and \erosita\ cluster surveys.

This paper is organized as follows. In the next section we describe the characteristics 
of the simulation set used for this work, the algorithms used to identify the haloes and 
our method for the light-cone reconstruction. In Section~\ref{s:sz} we describe the model 
used to associate the SZ signal to each halo and discuss our results on the effect of 
neutrino mass on number counts and SZ power spectrum. In Section~\ref{s:xray} we discuss 
the aspects connected with the X-ray properties of galaxy clusters and provide forecasts 
on the possible detections by \erosita. We summarize and draw our conclusions in 
Section~\ref{s:concl}. For the purpose of our work we will use two different cosmological 
models with different values of the Hubble parameter $h$. However in order to provide a 
consistent comparison we will fix $h=0.7$ ($h_{70}=1$) for both models whenever 
describing the predicted observational properties.

%%%%%%%%%%%%%%%%%%%%%%%%%%%%%%%%%%%%%%%%%%%%%%%%%%%%%%%%%%%%%%%%%%%%%%%%%%%%%%%%%%%%%%%%%%
%%%%%%%%%%%%%%%%%%%%%% Modelling the effect of neutrinos on the LSS %%%%%%%%%%%%%%%%%%%%%%
%%%%%%%%%%%%%%%%%%%%%%%%%%%%%%%%%%%%%%%%%%%%%%%%%%%%%%%%%%%%%%%%%%%%%%%%%%%%%%%%%%%%%%%%%%

\section{Modelling the effect of neutrinos on the LSS} \label{sec:models}

%%%%%%%%%%%%%%%%%%%%%%%%%%%%%%%%%%%%%%%%%%%%%%%%%%%%%%%%%%%%%%%%%%%%%%%%%%%%%%%%%%%%%%%%%%
\subsection{The simulation set} %%%%%%%%%%%%%%%%%%%%%%%%%%%%%%%%%%%%%%%%%%%%%%%%%%%%%%%%%%
For this work we use a set of cosmological simulations performed with the tree particle 
mesh-smoothed particle hydrodynamics (TreePM-SPH) code \gadgetthree, an evolution of the 
original code \gadget\ \citep{springel05} specifically modified by \cite{viel10} to 
account for the neutrino density evolution. Here we summarize the characteristics of the 
code and we refer to the original work for the details \citep[see also][]{bird12,
villaescusa13,villaescusa14}. The code follows the evolution of CDM and neutrino 
particles treating them as two separated collisionless fluids. Given the relatively 
higher velocity dispersion, neutrinos have a characteristic clustering scale larger than 
the CDM one, allowing to save computational time by neglecting the calculation of the 
short-range gravitational force. This results in a different spatial resolution for the 
two components: fixed by the PM grid for neutrinos and about one order of magnitude 
higher for CDM.

Starting from initial conditions with the same random phases, we run a total of six 
different cosmological simulations on very large scales, following a comoving volume of 
(2 \hmone\ Gpc)\cub\ from $z=99$ to present epoch, filled with 1024\cub\ dark matter 
particles and, where present, an equal amount of neutrino particles. No baryon physics is 
included in these simulations. We choose the cosmological parameters of the first three 
simulations, dubbed P0, P17 and P34, according to the \planck\ results 
\citep{planck14cp}, namely a flat \lcdm\ model which we generalize to a \lcdmn\ 
framework by changing only the value of the sum of the neutrino masses \smnut, 
respectively, and keeping fixed \omegam\ and the amplitude of primordial scalar 
perturbations $A_{\rm S}$. For the other simulation set (W0, W17 and W34) we assume the 
baseline \lcdm\ cosmology derived from the nine-years results of the \wmap\ satellite 
\citep{hinshaw13}, and introduce massive neutrinos with masses \smnut\ in the same way. 
For each simulation we produced 64 outputs logarithmically equispaced in the scale factor 
$a=1/(1+z)$, in the redshift interval $z=0-99$. The list of cosmological and numerical 
parameters assumed for the two simulation sets is reported in Table~\ref{t:cospar}.

\begin{table*}
\begin{center}
\caption{Set of parameters assumed in our six simulations. Second and third columns: 
density parameter of the cosmological constant and CDM, respectively, in per cent units. 
Fourth column: neutrino mass fraction, defined as \fnu$\equiv$
\omeganu/\omegam, in per cent units. The values of \omegal, \omegacdm\ 
and \omeganu\ are set to be consistent with a flat universe ($\Omega=1$). From fifth to 
eighth column, respectively: sum of the neutrinos masses (\smnu), Hubble constant 
(\hzero), normalization of the primordial power spectrum of the density fluctuations 
($A_{\rm S}$) and rms of matter density fluctuations in a sphere of 8\hmone Mpc (\sigmae) 
at $z$=0. The ninth column shows the number of particles used in the simulations 
($N_{\rm p}$) and the last two columns show the mass of the CDM and neutrino particles, 
respectively. For all simulations the comoving volume size is (2\hmone\ Gpc)\cub.}
\begin{tabular}{rcccccccccccc}
\hline
\hline
Simulation & & $10^2\,$\omegal & $10^2\,$\omegacdm & $10^2\,$\fnu & \smnu\ & 
\hzero\    &  $10^9 A_{\rm S}$  & \sigmae\ & 
& $N_{\rm p}$ & $m_{{\rm p}, \rm cdm}$       & $m_{{\rm p},\nu}$           \\
           & &                 &                   &                  & (eV) & 
(km s$^{-1}$ Mpc$^{-1}$) &               &  ($z$=0)            & 
&       & ($10^{10}$\hmone\msun) & ($10^{10}$\hmone\msun) \\
\hline
P0  && 68.39 & 31.61 &    0 & 0    & 67.11 & 2.15 & 0.834 && 
1024\cub & 65.36 & $-$  \\
P17 && 68.39 & 31.20 & 1.30 & 0.17 & 67.11 & 2.15 & 0.794 && 
2$\times$1024\cub & 64.53 & 0.84 \\
P34 && 68.39 & 30.80 & 2.56 & 0.34 & 67.11 & 2.15 & 0.751 && 
2$\times$1024\cub & 63.69 & 1.68 \vspace{0.12cm} \\
W0  && 72.90 & 27.10 & 0    & 0    & 70.3  & 2.43 & 0.809 && 
1024\cub & 56.04 & $-$  \\
W17 && 72.90 & 26.73 & 1.37 & 0.17 & 70.3  & 2.43 & 0.767 && 
2$\times$1024\cub & 55.28 & 0.76 \\
W34 && 72.90 & 26.36 & 2.73 & 0.34 & 70.3  & 2.43 & 0.723 && 
2$\times$1024\cub & 54.51 & 1.53 \\
\hline
\hline
\label{t:cospar}
\end{tabular}
\end{center}
\end{table*}

%%%%%%%%%%%%%%%%%%%%%%%%%%%%%%%%%%%%%%%%%%%%%%%%%%%%%%%%%%%%%%%%%%%%%%%%%%%%%%%%%%%%%%%%%%
\subsection{Halo identification} %%%%%%%%%%%%%%%%%%%%%%%%%%%%%%%%%%%%%%%%%%%%%%%%%%%%%%%%%
\label{ss:haloid}
First, we process the simulation outputs with the friends-of-friends (FoF) algorithm 
included in the \gadgetthree\ package. The code is applied only to CDM 
particles with linking length fixed to 0.16 times the mean 
interparticle distance. The minimum number of particles to identify a structure is set 
to 32, thus fixing its minimum mass to $M_{\rm FoF}\simeq2\times 10^{13}$\hmone\msun. 
Neutrino particles are subsequently attached to a given parent halo if the closest CDM 
particle belongs to it. The FoF catalogues are then processed with the \subfind\ 
algorithm \citep{springel01,dolag09} included in the \gadgetthree\ package, which 
identifies locally overdense gravitationally bound regions within an input parent halo. 
Here we adopt a minimum number of $20$ particles to make a valid sub-halo. Moreover, we 
use a specific routine included in the \subfind\ algorithm to compute spherical 
overdensities for FoF groups, and, in particular, for each halo we compute the value of 
\rfiveh, defined as the radius enclosing a matter (CDM+neutrinos) density equal to 500 
times the critical density of the Universe $\rho_{\rm c}(z)$ at the cluster redshift, and 
the corresponding mass in terms of \mfiveh$\equiv500\rho_{\rm c}(z)\times\frac{4}{3}\upi$
\rfiveh\cub. The choice of the density contrast $\delta=500$ is done in order to simplify 
the use of scaling relations to compute observables (see Sections~\ref{s:sz} and 
\ref{s:xray}). We stress also that the density threshold adopted in our study is 
significantly higher with respect to other works \cite[see, e.~g.][]{villaescusa14}: this 
makes the contribution of the neutrino component to the total mass of the sub-haloes 
negligible. With this procedure some of the initial FoF parent haloes are split into 
multiple sub-haloes, with the result of an increase of the total number of identified 
objects and of a lower minimum mass limit.

\begin{figure*}
\includegraphics[width=0.90\textwidth]{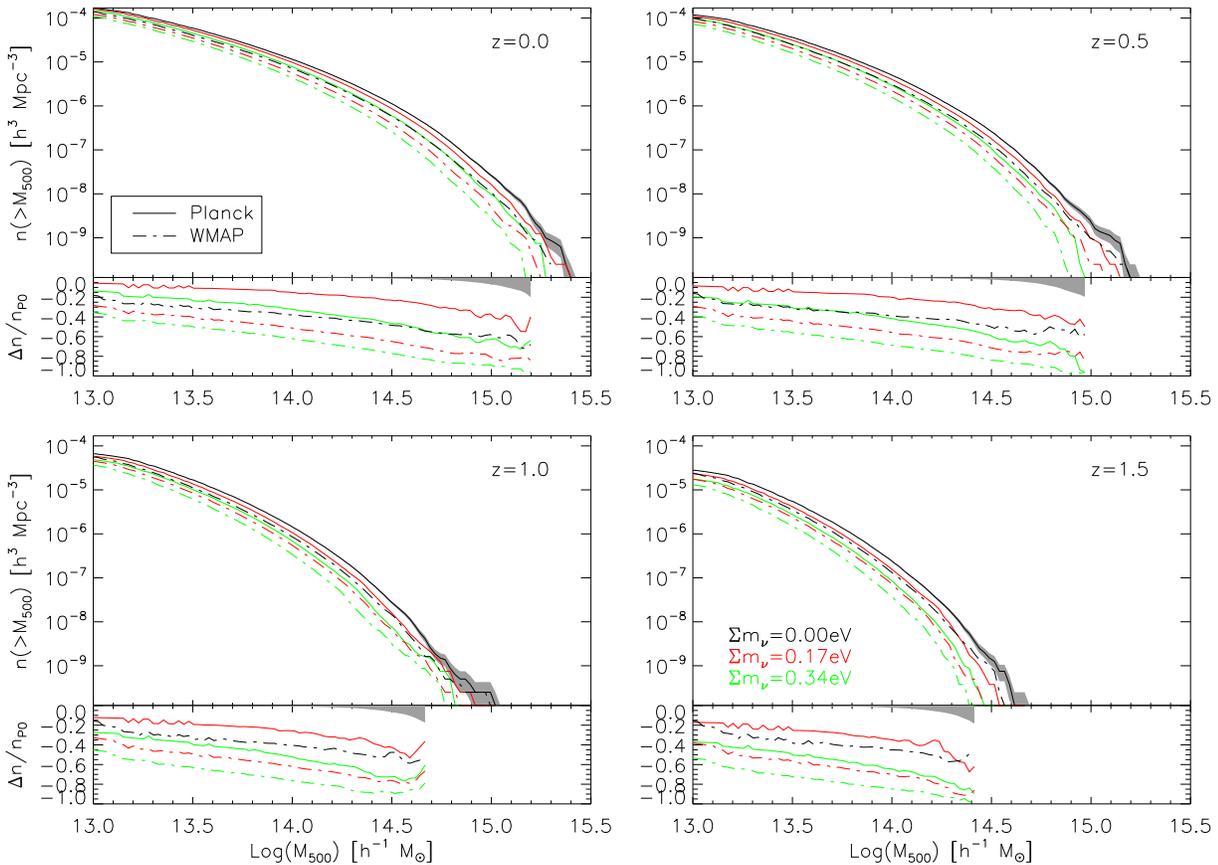}
\caption{
  Upper subpanels: mass function, in terms of comoving number density of haloes above a 
  given \mfiveh, for the two simulation sets at different redshifts. Solid lines refer 
  to the Planck simulation set (P0, P17, P34), dot-dashed lines refer to the WMAP 
  simulation set (W0,W17,W34). The three different values of \smnut\ are shown in black, 
  red and green, respectively. In the lower subpanels we show the fractional differences 
  $\Delta n/n$, with respect to the P0 simulation. The grey shaded area encloses the 
  Poissonian error computed over the simulation volume for the P0 model only.
  }
\label{f:mfun}
\end{figure*}

We show in Fig.~\ref{f:mfun} the mass functions of our six simulations in terms of number 
density of haloes above a given \mfiveh\ at four different redshifts. As expected, the 
presence of massive 
neutrinos slows down structure formation resulting in a smaller number of haloes per 
fixed mass at all redshifts. When looking at the $z=0$ plot, in the mass range \mfiveh=
10$^{13.5-14.5}$\hmone\msun\ the P17 model shows about 10--20 per cent less haloes than 
the P0 one, with differences doubled when considering the P34 and P0 models. Similar 
differences are found for the WMAP simulation set. We can also observe that the action of 
massive neutrinos is not completely degenerate with respect to the remaining cosmological
parameters if we take into account the redshift evolution and the shape of the mass 
function. More precisely, the decrease in the number of haloes due to neutrino 
free-streaming shows a steeper dependence on both mass and redshift.

\begin{figure*}
\includegraphics[width=0.90\textwidth]{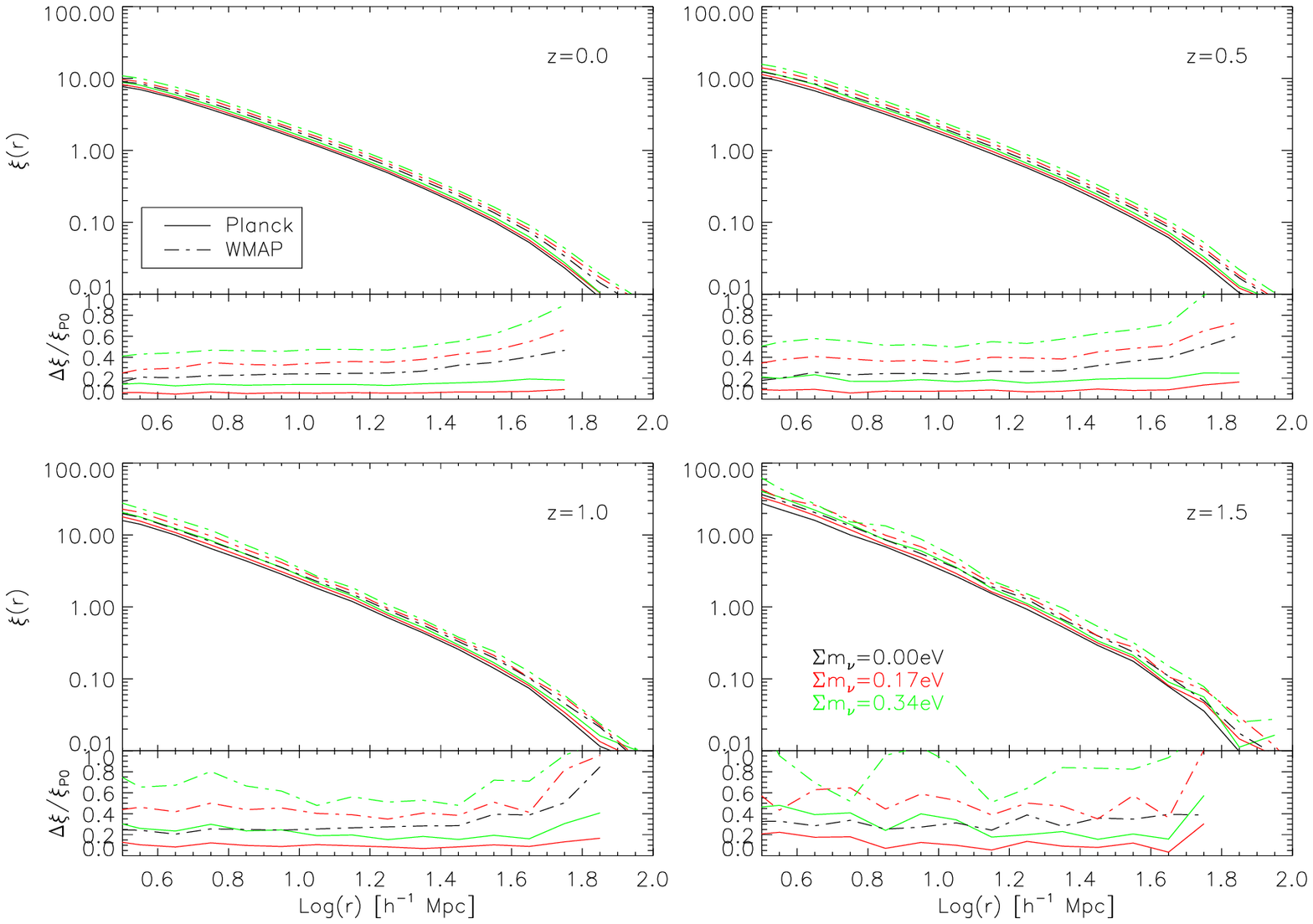}
\caption{
  Two-point spatial autocorrelation of haloes (\mfiveh$>10^{13.5}$\hmone\msun) as a 
  function of comoving distance for the two simulation sets at different redshifts. Lines 
  and colour coding are the same as in Fig.~\ref{f:mfun}. In the lower subpanels we show 
  the fractional differences $\Delta \xi/\xi$, with respect to the P0 simulation.
  }
\label{f:xi}
\end{figure*}

Massive neutrinos affect also the clustering of large-scale structures, as shown by the 
two-point spatial autocorrelation function of haloes in Fig.~\ref{f:xi}. Differently from 
the mass function case, massive neutrinos enhance halo clustering and the value of $\xi$ 
at all scales and redshifts \citep[see the discussion in][]{marulli11}. Considering the 
Planck simulation set, we observe that massive neutrinos produce a 20 per cent increase 
of the value of $\xi$ for the P34 simulation with respect to the P0 one, independent of 
scale and almost independent of redshift up to $z=1$. Similar differences are found in 
the WMAP simulation set. On the other side, the increase of the value of $\xi$ associated 
with the baseline cosmology is significantly higher: the W0 simulation shows a 20 per 
cent higher correlation than the P0 one at small scales, with differences increasing with 
distance up to 40 per cent at 50\hmone\ Mpc. This indicates that the use of cluster 
counts and cluster correlation function can help to break the degeneracy between neutrino 
mass and cosmological parameters \citep[see also the discussion in][]{sanchez14}.

%%%%%%%%%%%%%%%%%%%%%%%%%%%%%%%%%%%%%%%%%%%%%%%%%%%%%%%%%%%%%%%%%%%%%%%%%%%%%%%%%%%%%%%%%%
\subsection{Light-cone construction}%%%%%%%%%%%%%%%%%%%%%%%%%%%%%%%%%%%%%%%%%%%%%%%%%%%%%%
\label{ss:lcone}
The method to create the past light-cones from the simulation outputs is similar to the 
one described in our previous works \citep[see][]{roncarelli06a,roncarelli07,
roncarelli10a,roncarelli12}. The procedure consists in stacking the different simulation 
volumes along the line of sight, down to a redshift limit that we fix for our purposes to 
$z=3$, corresponding to 4381\hmone\ (4601\hmone) comoving Mpc for the \planck\ (\wmap) 
baseline cosmology, thus requiring to stack three times the simulation volume to fill up 
the light-cone. In order to exploit the full redshift sampling of the simulation outputs, 
each simulation volume is divided into slices along the line of sight to which we assign 
a different snapshot. The interval corresponding to the different slices is chosen by 
computing the age of the Universe as a function of the comoving distance from the 
observer and by assigning to each slice the snapshot that better matches this quantity. 
Given our set of snapshots we obtain 39 slices for both simulation sets.

In order to avoid the superimposition of the same structures along the line of sight, 
every simulation box undergoes a randomization process that consists of four steps: 
(i) we assign a 50 per cent probability to reflect each side of the output, (ii) we 
randomly choose the axis to be oriented along the line of sight and the direction of the 
other two, (iii) we perform a random recentring of the spatial coordinates by exploiting 
the periodic boundary conditions and (iv) we rotate the cube along the line of sight 
choosing a random angle. The slices belonging to the same cube volume undergo the same 
randomization process in order to preserve the large-scale power information. We repeat 
this process using the same random seeds for our six simulations: this, together with the 
use of the same phases in the initial conditions, ensures that we represent the same 
comoving volume with six different cosmological models, reducing the impact of cosmic 
variance on our results.

We fix the opening angle of each light-cone to 10$^\circ$ per side, a limit that ensures 
the validity of the flat-sky approximation, enclosing a comoving volume of 0.854 (0.977) 
$h^{-3}$ Gpc\cub\ for the Planck (WMAP) simulation set. By varying the initial random 
seed we generate 50 different light-cone realizations that we use to assess the 
statistical robustness of our results, thus covering a total area of 5000 deg\sqr. 
Despite the very large size of our simulation volume, this total area cannot be 
considered completely independent. In fact, the projected comoving size of the simulation 
at $z=3$ corresponds for the Planck (WMAP) model to 681 (622) deg\sqr, indicating that at 
the end of the light-cone we are replicating the same structure about 7 (8) times. 
However in the redshift range $0<z<0.6$, where most of the X-ray and SZ signal is 
expected, the simulation size covers all of the 5000 deg\sqr, allowing us to consider the 
comoving volume of our 50 light-cones as practically independent up to $z=0.6$.

%%%%%%%%%%%%%%%%%%%%%%%%%%%%%%%%%%%%%%%%%%%%%%%%%%%%%%%%%%%%%%%%%%%%%%%%%%%%%%%%%%%%%%%%%%
%%%%%%%%%%%%%%%%%%%%%%%%%%%%%%%%% Modelling the SZ signal %%%%%%%%%%%%%%%%%%%%%%%%%%%%%%%%
%%%%%%%%%%%%%%%%%%%%%%%%%%%%%%%%%%%%%%%%%%%%%%%%%%%%%%%%%%%%%%%%%%%%%%%%%%%%%%%%%%%%%%%%%%

\section{Modelling the SZ signal} \label{s:sz}

\begin{figure*}
\includegraphics[width=0.90\textwidth]{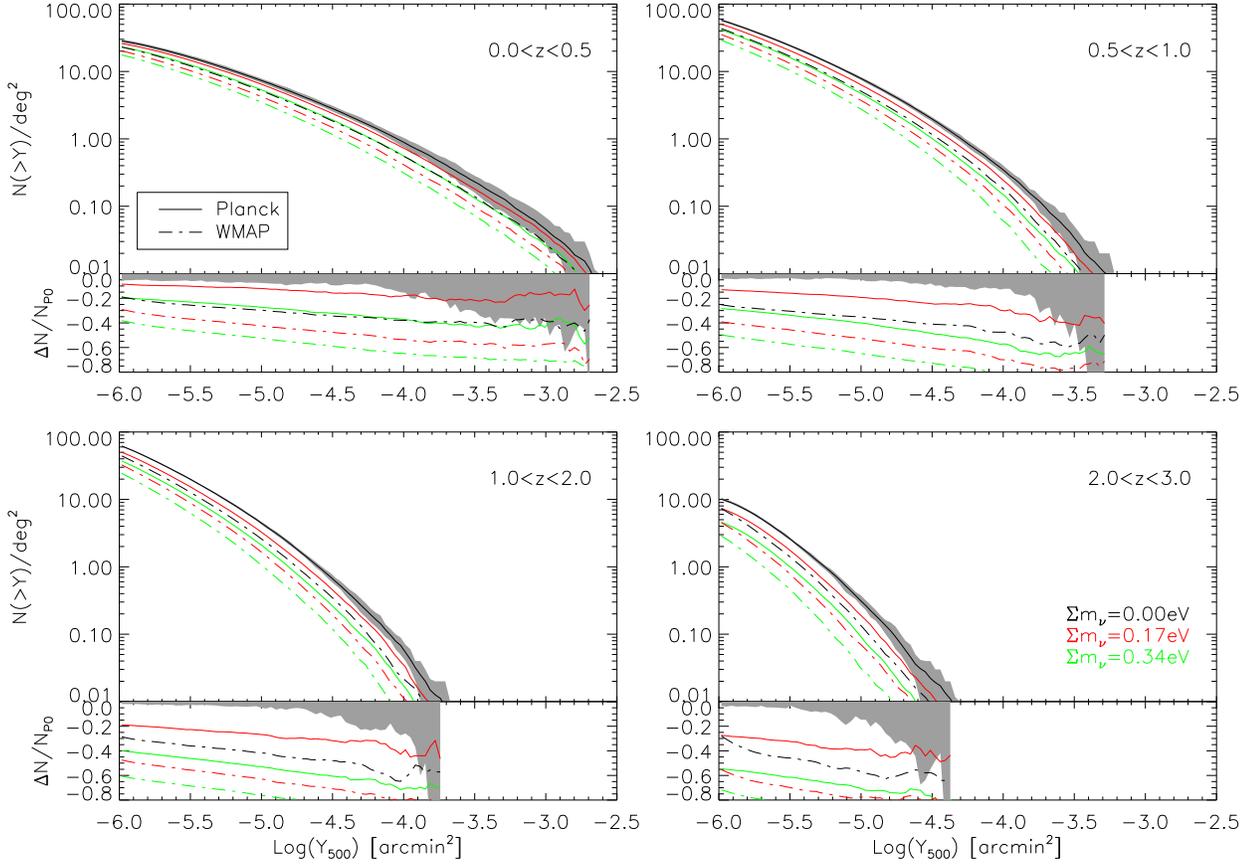}
\caption{
  Number of haloes per square degree as a function of the limit in \yfiveh. Lines and 
  colour coding are the same as in Fig.~\ref{f:mfun}, the grey shaded area encloses the 
  central 68 per cent of the values computed in the 50 different light-cones (100 
  deg\sqr\ each) for the P0 model.
  }
\label{f:ny}
\end{figure*}

Starting from the simulation outputs, we associate to each halo a value of the integrated 
Compton \ypar\ \yfiveh\ as a function of \mfiveh\ and $z$, using the scaling-law adopted 
by \cite{planck14cc}:
\begin{equation}
\frac{Y_{500}}{{\rm arcmin}^2} = 
E^\beta(z) \, \frac{10^{-4}\,Y_*}{6^{\alpha}} \,
\left(\frac{h_{70}^{-1} \, {\rm Mpc}}{d_{\rm A}(z)}\right)^{2}
\left[\frac{(1-b)M_{500}}{10^{14}\, h_{70}^{-1}\, M_{\sun}}\right]^\alpha \, ,
\label{e:ym}
\end{equation}
where $E(z) \equiv H(z)/H_0 = \sqrt{\Omega_{\rm m}(1+z)^3+\Omega_\Lambda}$, $\beta$ is 
fixed to the self-similar evolution value of 2/3 and $d_{\rm A}(z)$ is the angular 
diameter distance. The quantities $Y_*$, $\alpha$ and $b$ (normalization, slope and 
pressure bias, respectively) are taken from the baseline values of \cite{planck14cc} and 
fixed to 0.646, 1.79 and 0.2, respectively, for all haloes. Once the expected value of 
$Y_{500}$ is computed, the final value is determined by adding randomly an intrinsic 
scatter with a logarithmic standard deviation $\sigma_{\Log Y}=0.075$.

Under these assumptions we compute the expected number of haloes above a given \yfiveh\ 
in different redshift bins, and show the results in Fig.~\ref{f:ny}. The differences are 
comparable to what observed in Fig.~\ref{f:mfun} for the mass function, with decrements 
in the expected number counts roughly proportional to the value of \smnu. We can see, 
however, that while at low redshift the W0 model appears almost degenerate with the P34 
one, when moving towards higher redshift the effect of neutrinos is stronger. The cosmic 
variance (grey shaded area) computed as the interval that encloses the central 68 per 
cent of the values in the 50 different fields, is smaller than the difference between the 
various models for \yfiveh$\lesssim10^{-4}$ arcmin\sqr\ and approximately follows 
Poissonian statistics.

\begin{figure}
\includegraphics[width=0.5\textwidth]{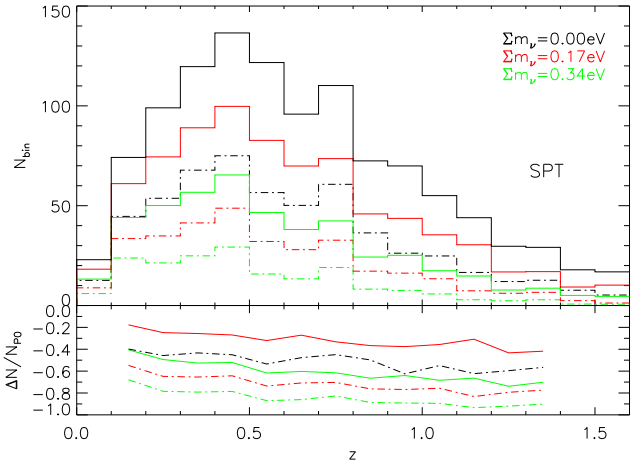}
\caption{
  Predicted number counts in different redshift bins ($\Delta z =0.1$) of galaxy clusters 
  detectable by \spt\ (assuming S/N$>4.5$ and a survey area of 2500 deg\sqr) for the two 
  sets of simulations described in the text. Lines and colour coding are the same as in 
  Fig.~\ref{f:mfun}. In the lower panel we show the fractional difference computed with 
  respect to the P0 simulation.
  }
\label{f:ndz_y}
\end{figure}

We use our models also to predict the expected detections by the \spt\ and \planck\ 
surveys. In general, the probability of detecting a galaxy cluster via the SZ effect 
depends not only on its \yfiveh\ value but also on the sky surface occupied by the 
halo: the latter determines the amount of noise, mainly primary CMB, and grows 
approximately linearly with the halo angular size. From the simulated light-cone 
catalogues we derive the value of $\theta_{500}=\arcsin(R_{500}/d_{\rm A})$ for our 
objects. Then by using simulated data of \spt\ noise (Saro \& Liu, private communication) 
we compute a value of $\Delta Y$ that we use as a noise estimate and fix a 
signal-to-noise ratio (S/N) detection threshold of 4.5 \citep[see][]{reichardt13}. 
Finally, we associate to each halo a detection probability by taking into account the 
intrinsic scatter in the \yfiveh--\mfiveh\ relation of equation~(\ref{e:ym}). We show in 
Fig.~\ref{f:ndz_y} the expected number of detections for the full \spt\ cluster survey 
(area of 2500 deg\sqr). We find that a value of \smnu=0.17 eV reduces the expected 
detections by 20 per cent at $z=0$ and up to 40 per cent at $z>1$ with respect to the 
\lcdm\ scenario in the \planck\ baseline cosmology. We can also see that for all models 
15--20 per cent of the total detections will be at $z>1$, where the neutrinos produce the 
largest differences. However the effect of neutrinos is almost degenerate with cosmology: 
in fact, the P34 and W0 models produce very similar results. This indicates that the 
\spt\ cluster survey alone will not be able to break the degeneracy between \smnu\ and 
\sigmae, and that a combination with data from other probes will be required to measure 
\smnu\ precisely.

\begin{figure}
\includegraphics[width=0.5\textwidth]{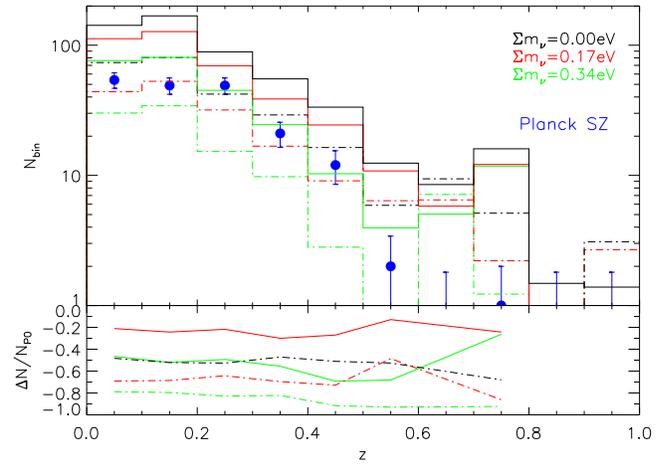}
\caption{
  Same as Fig.~\ref{f:ndz_y} but for galaxy clusters detectable by \planck\ (see text for 
  details). The blue points represent the detections by \protect \cite{planck14cc} with 
  1$\sigma$ error bars.
  }
\label{f:ndz_pl}
\end{figure}

To obtain the estimate for the \planck\ cluster survey with an equivalent procedure we 
would need a precise and complete instrumental noise information which is not publicly 
available at the moment. For this reason we proceed in the following way. We consider the 
mass selection functions, $M_{\rm lim}(z)$, published by \citeauthor{planck14cc} 
(\citeyear{planck14cc}, see their fig.~3) for the shallow, medium and deep surveys. Then 
for every halo in our catalogue we compute a detection probability $\chi(M_{500},z)$. 
Since this quantity actually depends on \yfiveh, which is related to \mfiveh\ with an 
intrinsic logarithmic scatter $\sigma_{\Log Y}$, we consider this effect by computing it 
in the following way:
\begin{equation}
\chi(M_{500},z) = 
\sum_{i=1}^{3}
\frac{f_i}{2}\left[1+\erf\left(
\frac{\Log\left(\frac{M_{500}}{M_{{\rm lim},i}(z)}\right)}{\sqrt{2}\sigma_{\Log Y}}
\right)
\right] \, ,
\label{e:chi}
\end{equation}
being $f_i$ the area fraction of the three surveys (48.7, 47.8 and 3.5 per cent for the 
shallow, medium and deep surveys, respectively) and $M_{{\rm lim},i}(z)$ the relative 
selection function. Finally, for each cluster in our light-cones we sum their individual 
detection probability and compute the expected number of detections per redshift bin 
considering a coverage area of 26 814 deg\sqr\ (i.e. the 65 per cent of the full sky). 
We note, however, that our prediction for the P0 model is higher by about 30 per cent 
with respect to the equivalent one in \cite{planck14cc}: we verified that this translates 
in a 10 per cent difference in the mass limit and is likely due to the approximations in 
our modelling of the \planck\ selection function. To overcome this problem, we apply this 
correction factor to the value of $M_{{\rm lim},i}(z)$ as it appears in 
equation~\ref{e:chi}. The corresponding results in terms of clusters number counts are 
plotted in Fig.~\ref{f:ndz_pl}. All the models that adopt \planck\ cosmology show an 
excess in the first two bins with respect to the real \planck\ detections (blue points), 
while at $z>0.2$ the P34 model is in agreement with the data: this is consistent with 
what found by \cite{planck14cc} who obtain a value of \smnu$=(0.53\pm0.19)$ when 
considering CMB and clusters together. However this appears to be somewhat in tension 
with the result of \smnu$=(0.20\pm0.09)$ obtained when adding BAO data. Again, we observe 
that with respect to number counts the P34 and W0 model are almost perfectly degenerate.

We now analyse our predictions for the average \ypar\ value and SZ power spectrum. By 
using our light-cone catalogues, and under the assumption of clusters' spherical symmetry 
and of a fixed pressure profile, we are able to produce a set of mock \ypar\ maps for 
each light-cone realization and for each simulation, and use them to predict global 
SZ properties. Starting from the position of the cluster centres in the light-cones, we 
assign to each object a projected pressure profile using a modification of the classic 
$\beta$-model \citep{cavaliere78}, namely a rolling-$\beta$ polytropic profile that 
allows for a steepening of the density and temperature profiles in the outskirts 
\citep[see][and discussions therein]{ameglio06,roncarelli06b,roncarelli10a}. The profile, 
expressed in terms of the \ypar, has the following form:
\begin{equation}
y(\theta)=
\left[1+\left(\frac{\theta}{\theta_{\rm c}}\right)^2\right]^
{(1-3\,\gamma\,\beta_{\rm eff})/2} \, ,
\end{equation}
where $\theta_{\rm c}$ is the angular size of the core radius, $\gamma$ is the polytropic 
index of the gas, fixed to 1.18, and $\beta_{\rm eff}$ is the effective slope, defined as
\begin{equation}
\beta_{\rm eff}\equiv
\frac{\beta_{\rm int}+\left(\frac{\theta}{\theta_{\rm c}}\right)\beta_{\rm ext}}
{1+\frac{\theta}{\theta_{\rm c}}} \, ,
\end{equation}
where $\beta_{\rm int}$ and $\beta_{\rm ext}$ represent the internal and external slopes 
of the profile, respectively. This method has the advantage of providing directly the 
projected quantity to map, with three free parameters, namely $\beta_{\rm int}$, 
$\beta_{\rm ext}$ and the ratio $\theta_{\rm c}/\theta_{500}$, that can be tuned to 
reproduce a given pressure profile. In particular we find the parameters values 
($\beta_{\rm int}=0.98$, $\beta_{\rm ext}=1.28$ and $\theta_{\rm c}/\theta_{500}=0.426$) 
that match the projection of the universal pressure profile of \cite{arnaud10} and apply 
them to our clusters. Given the high pressure contribution coming from external regions, 
we map the structure of our haloes up to 8~\rfiveh.

\begin{figure*}
\includegraphics[width=1.0\textwidth]{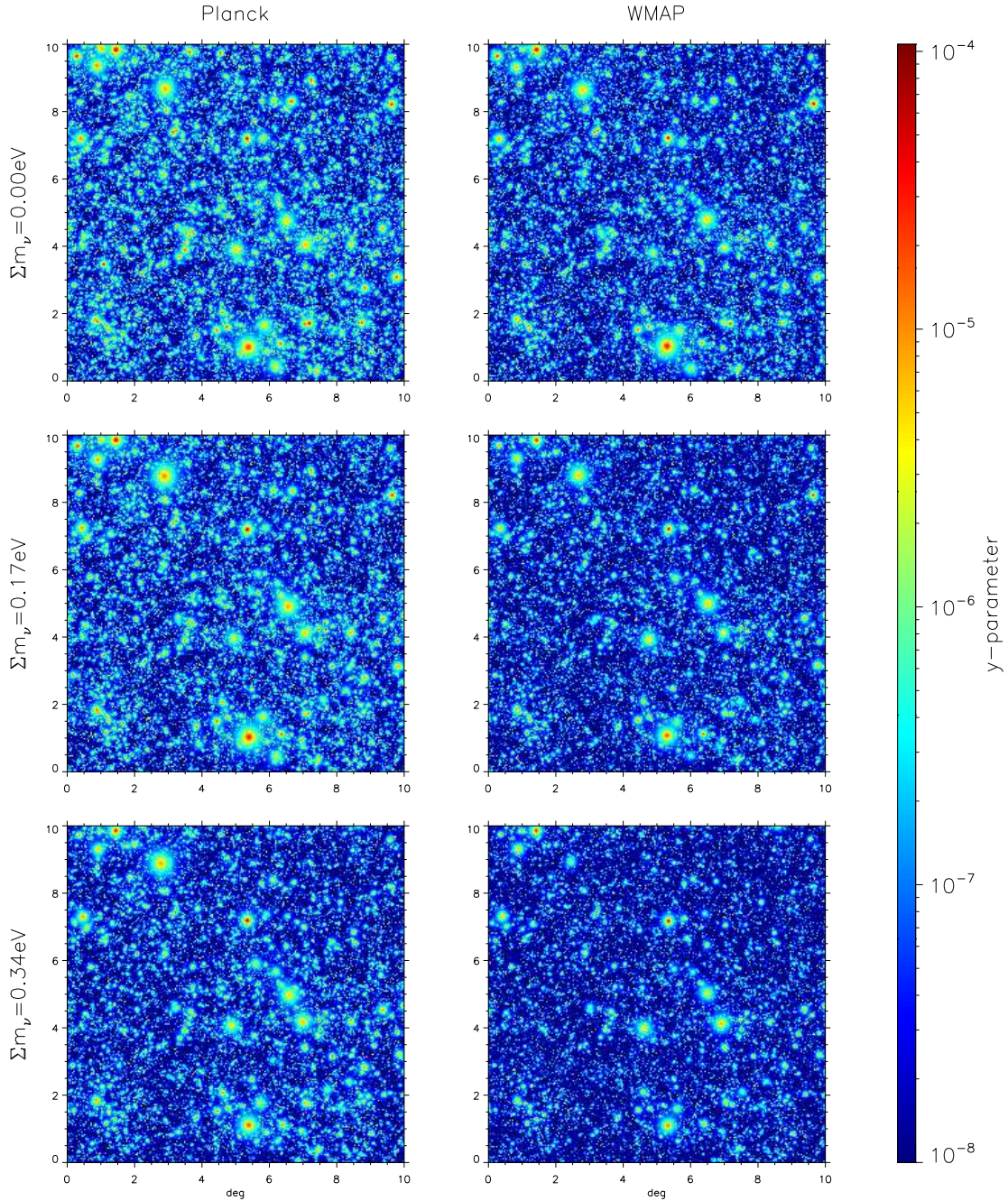}
\caption{
  Maps of the \ypar\ for the same light-cone in our six models. Left- and right-hand 
  columns correspond to the Planck and WMAP simulation set, respectively. Top, middle and 
  bottom rows show the maps for \smnut, respectively. Each map is 10$^\circ$ per side and 
  considers all haloes with $z>0.03$. All maps adopt the same colour scale, displayed on 
  the right.
  }
\label{f:ymaps}
\end{figure*}

We show in Fig.~\ref{f:ymaps} the maps of the \ypar\ obtained with this method for the 
same light-cone realizations produced with the six simulations described in 
Table~\ref{t:cospar}. These maps represent the whole 100 deg\sqr\ sky patch after 
cutting out haloes whose centre is located at $z<0.03$. We note that, even reproducing 
the same light-cone volume in the six simulations (see Section~\ref{ss:lcone}), a 
significant cosmic variance from field to field is present, mainly due to the intrinsic 
scatter in the \yfiveh--\mfiveh\ relation. It is clear from these images that the 
diminished abundance translates into a lower Comptonization value with increasing 
neutrino mass and moving from \planck\ to \wmap\ cosmology. We show in Table~\ref{t:avy} 
the average \ypar\ computed over the whole 5000 deg\sqr. The quoted error represents the 
1$\sigma$ deviation of the 50 different fields, i.e. the range that encloses the central 
68 per cent of the values. First of all we must consider that these values are low when 
compared to other estimates obtained from hydrodynamical simulations, due to the lack of 
the diffuse component (i.e. filaments) which is expected to provide about half of the 
contribution to the whole integrated value \citep[see the discussion in][]{roncarelli07}: 
this means that we have to consider these results as representative of the fraction of 
the expected total value associated with galaxy clusters only. As expected, massive 
neutrinos reduce consistently the average \ypar, with P17 and P34 simulations having 
values about 20 and 40 per cent lower, respectively, with respect to the P0 model. When 
considering the WMAP simulation set, the reduction is even higher (25 and 45 per cent for 
W17 and W34, respectively), due to the higher value of the ratio \fnu. Considering the 
whole set of six simulations we observe that the average \ypar\ due to galaxy clusters 
scales as
\begin{equation}
\frac{\langle\,y_\nu\,\rangle}{\langle\,y_0\,\rangle} \ \approx 
\ \left(1-f_\nu\right)^{20} \, .
\end{equation}

\begin{table}
\begin{center}
  \caption{
  Average \ypar\ in the whole simulation set. The quoted error encloses the central 68 
  per cent of the values of the 50 light-cones. These values do not account for the 
  diffuse component, e.g. the gas outside galaxy clusters.
  }
\begin{tabular}{rcc}
\hline
\hline
Simulation  && $\langle\,y\,\rangle \, (10^{-7})$ \\
\hline
P0 && $4.82_{-0.37}^{+0.45}$ \\
P17 && $3.92_{-0.29}^{+0.32}$ \\
P34 && $2.95_{-0.19}^{+0.20}$ \vspace{0.12cm} \\
W0 && $3.23_{-0.21}^{+0.25}$ \\
W17 && $2.41_{-0.17}^{+0.17}$ \\
W34 && $1.84_{-0.14}^{+0.15}$ \\
\hline
\hline
\label{t:avy}
\end{tabular}
\end{center}
\end{table}

Another important quantity that can be used to observe the effect of neutrinos on the 
LSS of the Universe is the SZ power spectrum, which is known to be the dominant source of 
CMB anisotropies at small scales ($\lesssim 5$ arcmin) and that has been recently 
measured by \planck\ \citep{planck14ym}, \spt\ \citep{reichardt12,crawford14} and Atacama 
Cosmology Telescope \citep[\act;][]{sievers13}. Using the complete set of \ypar\ maps (6 
maps like the ones of Fig.~\ref{f:ymaps} for each light-cone realization) we compute 
their power spectrum with a method based on fast Fourier transform and in the flat-sky 
approximation, and express it in terms of power as a function of the multipole $\ell$.

\begin{figure*}
\includegraphics[width=1.0\textwidth]{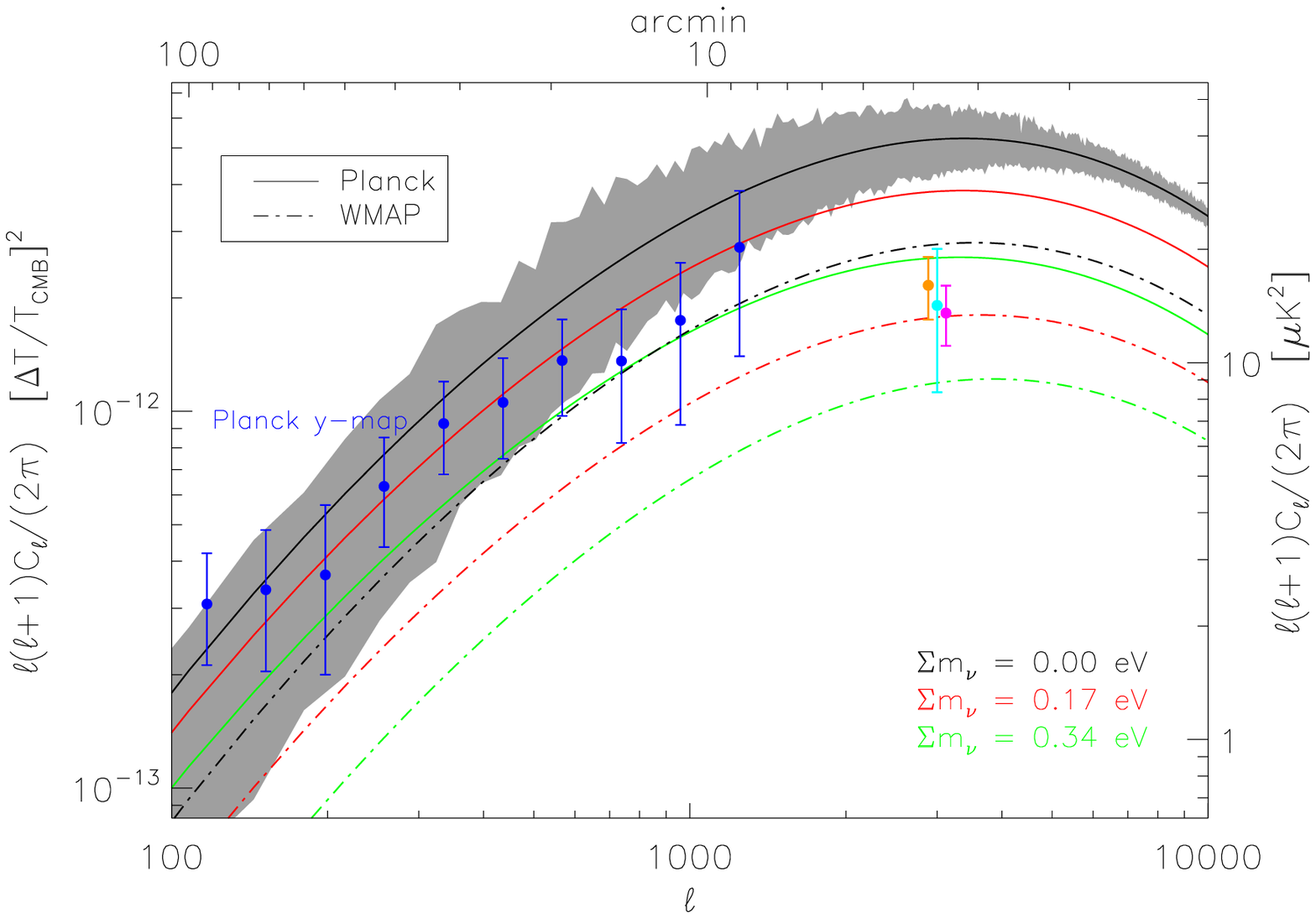}
\caption{
  Power spectrum of the SZ signal in the RJ limit as a function of the multipole $\ell$ 
  for the two simulation sets. Lines and colour coding are the same as in 
  Fig.~\ref{f:mfun}. All these lines represent the average over the 50 light cone 
  realizations, while the grey shaded area encloses the 1$\sigma$ deviation between the 
  50 light-cones for the P0 simulation only (see text for details). The blue points 
  represent the SZ power spectrum measurements by \planck\ \protect\citep{planck14ym}. 
  The orange and magenta points represent the \spt\ results by \protect 
  \cite{reichardt12} and \protect\cite{crawford14}, respectively. The cyan point 
  represents the \act\ measurement by \protect \cite{sievers13}.
  }
\label{f:powsp}
\end{figure*}

Since for each  model the power spectrum shows significant changes from one light-cone to 
another due to the cosmic variance, in order to define a global quantity for each 
simulation we proceed in the following way. For every simulation and for each value of 
the multipole $\ell$ we consider the values of $P(\ell)\equiv\ell(\ell+1)C_\ell/(2\upi)$ 
obtained in the 50 different light-cones and, after defining $x\equiv {\rm Log} 
(\ell)$, we compute the average $P(x)$ for each model and we quantify their dispersion 
by considering the eight highest and eight lowest values and by defining $\sigma(\ell)$ 
as the half of their difference. Then, we observe that the result can be fit with a 
skew-normal distribution \citep{azzalini85} with shape parameter fixed\footnote{
This value is chosen to simplify the expression in equation~(\ref{e:sknorm}) and 
corresponds to a skewed Gaussian curve, with skewness equal to $-0.272$.}
to $\alpha=-\sqrt{2}$, thus with the following formula:
\begin{equation}
P(x) = \frac{A}{\sqrt{2\upi}\,\sigma} \, \exp\left({-\frac{(x-\mu)^2}{2\sigma^2}}\right) 
       \left[1+\erf\left(\frac{\mu-x}{\sigma}\right)\right] \, ,
\label{e:sknorm}
\end{equation}
where $A$, $\mu$ and $\sigma$ are left as free parameters. Finally, for each model we 
fit the average power spectrum by using the value of $\sigma(\ell)$ as an error for each 
point. We report in Table~\ref{t:psfit} the best-fitting parameters for each model and 
plot in Fig.~\ref{f:powsp} the corresponding SZ power spectra in the Rayleigh-Jeans (RJ) 
limit. For all the models the peak position, defined by the $\mu$ parameters show small 
changes. On the other hand, the \wmap\ models show slightly lower values of $\sigma$ and, 
most importantly, the normalization is highly influenced by the presence of massive 
neutrinos: the P17 and W17 models show a reduction of 25-30 per cent with respect to 
the corresponding \lcdm\ model, while for the P34 and W34 models the normalization is 
reduced to less than a half. We find that its dependence on 
\fnu\ scales approximately as
\begin{equation}
\frac{A_\nu}{A_0} \ \approx \ \left(1-f_\nu\right)^{25-30} .
\end{equation}

\begin{table}
\begin{center}
  \caption{
  Properties of the SZ power spectrum for our six simulations. From second to fourth 
  column: values of the best-fitting parameters (see text for details) to the simulated 
  data using equation~(\ref{e:sknorm}). The fifth column indicates the value of the 
  best-fitting function at $\ell=3000$, with the errors that represent the interval 
  enclosing the central 68 per cent of the values of the 50 different light-cone 
  realizations. The values $A$ and $A_{3000}$ are expressed in terms of 
  $(\Delta T/T_{\rm CMB})^2$ in the RJ limit.
  }
\begin{tabular}{rccccccc}
\hline
\hline
Simulation  &  & $A\,(10^{-12})$ & $\mu$ & $\sigma$  & & $A_{3000}\,(10^{-12})$ \\
\hline
P0 & & 6.84 & 3.90 & 0.692 & & $5.27_{-1.01}^{+1.16}$ \\
P17 & & 5.03 & 3.91 & 0.700 & & $3.84_{-0.68}^{+0.75}$ \\
P34 & & 3.38 & 3.90 & 0.706 & & $2.55_{-0.54}^{+0.44}$ \vspace{0.12cm} \\
W0 & & 3.58 & 3.92 & 0.685 & & $2.77_{-0.53}^{+0.54}$ \\
W17 & & 2.33 & 3.93 & 0.693 & & $1.78_{-0.35}^{+0.38}$ \\
W34 & & 1.54 & 3.95 & 0.678 & & $1.19_{-0.24}^{+0.23}$ \\
\hline
\hline
\label{t:psfit}
\end{tabular}
\end{center}
\end{table}

Our results can be compared to the measurements at $\ell<1000$ of \cite{planck14ym} 
and to the ones at $\ell=3000$ of \spt\ \citep{reichardt12,crawford14} and \act\ 
\citep{sievers13}, also plotted in Fig.~\ref{f:powsp}. As a reference, we also show in 
Table~\ref{t:psfit} the value of $A_{3000}$ for our six models in the RJ limit
\footnote{To obtain the corresponding values at 150 GHz (\spt) and 148 GHz (\act) one 
should divide our results by a factor of 4.40 and 4.17, respectively.}.
We can see that none of our models is able to reproduce the flattening obtained by  
\planck\ at large-scales, resulting in a significant underestimate of the measured 
spectrum at $\ell\lesssim 200$: this is likely due to the lack in our model of the 
diffuse component associated with non-virialized structures. The points at 
$\ell\gtrsim200$ are well fitted by our P17 model, with also the P34 and W0 providing 
acceptable fits to the data. Considering the uncertainty due to the cosmic variance, this 
is in broad agreement with the result of \cite{planck14ym}, who obtain best-fitting 
values of \sigmae$ = 0.74 \pm 0.06$ and \omegam$ = 0.33 \pm 0.06$. On the other hand, the 
results at $\ell=3000$ favour the P34, W0 and W17 models which are all consistent at 
about 1$\sigma$. The P17 model shows an excess at more than 3$\sigma$ with respect to 
both \spt\ measurements and at 2$\sigma$ with respect to \act\ data: this highlights a 
general difficulty of this kind of models fitting both low and high $\ell$ measurements 
of the SZ power spectrum.

%%%%%%%%%%%%%%%%%%%%%%%%%%%%%%%%%%%%%%%%%%%%%%%%%%%%%%%%%%%%%%%%%%%%%%%%%%%%%%%%%%%%%%%%%%
%%%%%%%%%%%%%%%%%%%%%%%%%%%%%%%%% Modelling X-ray counts %%%%%%%%%%%%%%%%%%%%%%%%%%%%%%%%%
%%%%%%%%%%%%%%%%%%%%%%%%%%%%%%%%%%%%%%%%%%%%%%%%%%%%%%%%%%%%%%%%%%%%%%%%%%%%%%%%%%%%%%%%%%

\section{Modelling X-ray counts} \label{s:xray}
In this section we aim at describing the X-ray properties of the galaxy clusters in our 
simulations and predicting the expected number counts for the \erosita\ all-sky survey, 
that will constitute an important probe of the LSS in the next future. 
However, this kind of modelling is significantly more difficult than the 
corresponding SZ one, mainly due to the uncertainties associated with the X-ray scaling 
laws and to their redshift evolution, making our results uncertain in terms of absolute 
values. For this reason, we focus on the discussion of the relative differences 
associated with the presence of massive neutrinos, that are less dependent on our model 
assumptions.

In order to assign to each galaxy cluster an X-ray luminosity we adopt the scaling-law of 
\cite{mantz10a}, namely
\begin{equation}
\frac{L_{500}}{10^{44}\,h_{70}^{-2}\,{\rm erg/s}} = 
E(z)10^{\beta_0} 
\left(\frac{E(z)\,M_{500}}{10^{15}\,h_{70}^{-1}\,M_{\sun}}\right)^{\beta_1} \, ,
\label{e:lm}
\end{equation}
where $L_{500}$ is the bolometric X-ray luminosity inside \rfiveh, and the two parameters 
$\beta_0$ and $\beta_1$ are fixed to their best-fitting values of 1.23 and 1.63, 
respectively. Here we are assuming a self-similar redshift evolution. After computing the 
expected value of $L_{500}$ we also apply an intrinsic logarithmic scatter characterized 
by $\sigma_{\Log L}=0.17$. We then convert each bolometric luminosity to the 
corresponding flux in the 0.5--2 keV band
\begin{equation}
S_{[0.5-2]}=\frac{L_{500}\,f(T_{500},z)}{4\upi\,d_{\rm L}^2(z)} \, ,
\end{equation}
where $d_{\rm L}(z)$ is the luminosity distance, $f(T_{500},z)$ is the band correction, 
computed assuming a bremsstrahlung emission. The cluster temperature $T_{500}$ is also 
computed applying the scaling-law obtained by \cite{mantz10a} and assuming a self-similar 
redshift evolution.

Under these assumptions we compute the number of clusters detectable by the \erosita\ 
full-sky survey by adopting a simple flux threshold of $3 \times 10^{-14}$ \flunits\ in 
the 0.5--2 keV band, as reported in \cite{merloni12}. We stress that the problem of 
determining the actual detectability of a galaxy cluster in the real survey is a more 
complicated issue, connected to the capability of disentangling faint cluster candidates 
from contaminating sources (mainly faint active galactic nuclei at high redshift) given 
the large point spread function of \erosita\ \citep[see e.g. a discussion on this problem 
with {\it XMM-Newton} observations in][]{brusa10}, which is beyond the scope of this 
paper. For this reason our results should be interpreted as upper limits of the \erosita\ 
survey capability.

\begin{figure*}
\includegraphics[width=0.90\textwidth]{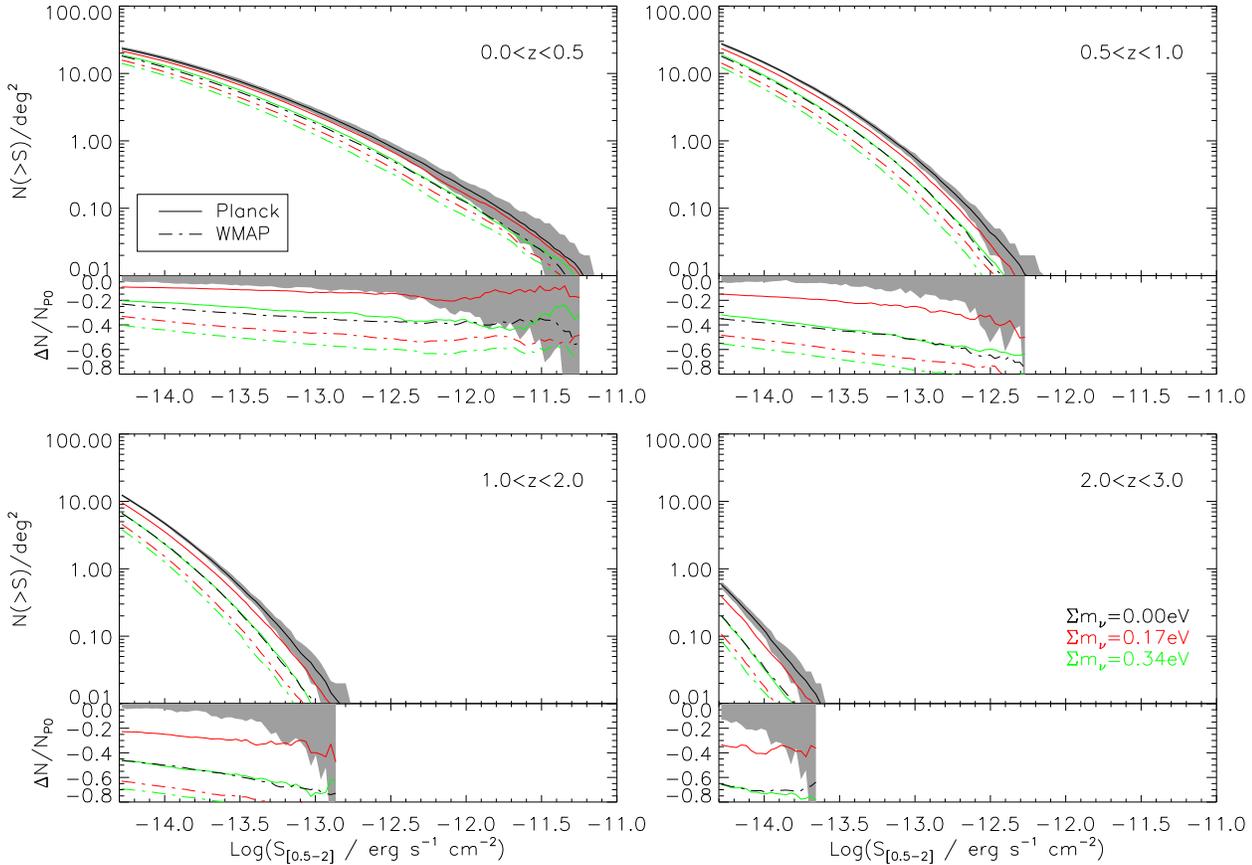}
\caption{
  Same as Fig.~\ref{f:ny}, but for the number of galaxy clusters as a function of limit 
  flux in the 0.5--2 keV band.
  }
\label{f:ns}
\end{figure*}

We show in Fig.~\ref{f:ns} the log$N$-log$S$ of the haloes in our 50 light-cones. These 
results look similar with what already said for Fig.~\ref{f:ny}, with the inclusion of 
massive neutrinos causing comparable decrements in the expected number of haloes. In this 
case the different redshift evolution of the X-ray flux makes the degeneracy with respect 
to cosmology more severe, especially between the P34 and W0 models at all redshifts. This 
is mainly due to the value of the \omegam\ parameter that influences the redshift 
evolution through $E(z)$ (see equation~\ref{e:lm}).

\begin{figure}
\includegraphics[width=0.5\textwidth]{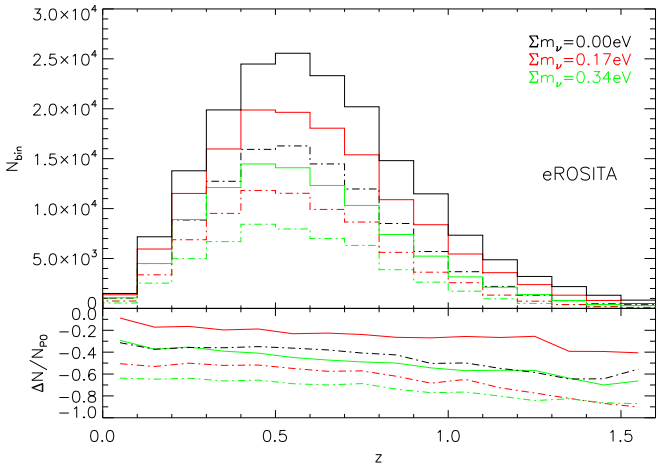}
\caption{
  Predicted number counts in different redshift bins ($\Delta z =0.1$) of galaxy clusters 
  (\mfiveh$> 10^{14}$\hmone\msun) detectable by the \erosita\ all-sky survey 
  ($S_{[0.5-2]} > 3 \times 10^{-14}$ \flunits) for our two sets of simulations. Lines and 
  colour coding are the same as in Fig.~\ref{f:mfun}. In the lower subpanel we show the 
  fractional difference computed with respect to the P0 simulation.
  }
\label{f:ndz_s}
\end{figure}

The potential of the \erosita\ survey comes out more clearly when analysing the expected 
detections in the redshift bins shown in Fig.~\ref{f:ndz_s}. For this computation we 
considered only haloes with \mfiveh$>10^{14}$\hmone\msun, in order to avoid further 
uncertainties in the scaling relations adopted. With our assumptions the \erosita\ 
cluster survey would be almost volume limited for the haloes above this mass. The total 
number of potential detections for the P0 model is about 180 000, which is consistent 
with what quoted by \cite{merloni12}. This makes the differences associated with a change  
in the value of \smnu\ highly significant in terms of the corresponding Poissonian 
uncertainties in all redshift bins.

\begin{figure}
\includegraphics[width=0.5\textwidth]{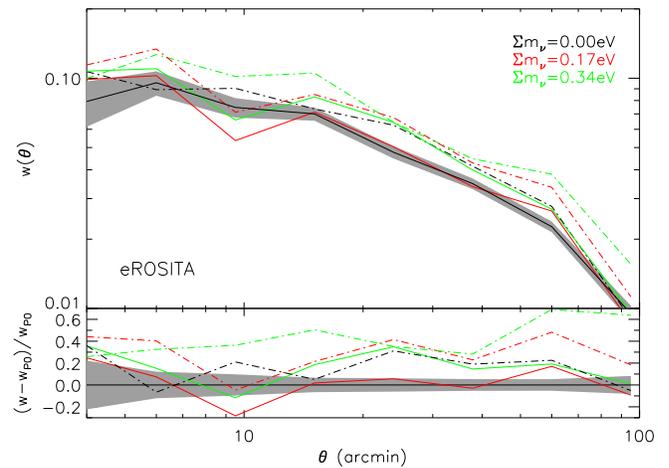}
\caption{
  Correlation function as a function of the angular separation for the haloes detectable 
  by the \erosita\ cluster survey (see text for details). The grey-shaded area shows the 
  Poissonian error (1$\sigma$) for the P0 simulation only computed over the 50 
  light-cones (5000 deg\sqr). In the lower subpanel we show the fractional difference 
  computed with respect to the P0 simulation.
  }
\label{f:xiang}
\end{figure}

Given the very high number of potential detections we can compute the corresponding 
angular correlation function $w(\theta)$ for the six models considered in this work and 
check if it can break the degeneracy between \smnu\ and cosmology, as said in 
Section~\ref{ss:haloid}. The results plotted in Fig.~\ref{f:xiang} show a trend similar 
to what happens for the cluster correlation function shown in Fig.~\ref{f:xi}, with 
a higher correlation for \wmap\ cosmology and for higher neutrino masses, with 
differences that are larger than the typical Poissonian error at $\theta \gtrsim 10$ 
arcmin. However, even if the fractional differences are high, the value of $w(\theta)$ at 
these angular scales is small (below 0.1 for all models), indicating that the systematics 
associated with the scaling law, and in particular to the redshift evolution, will 
dominate the uncertainty making it difficult to disentangle between the different models. 
This indicates that in order to exploit the correlation function a good control of 
observational systematics is required, together with precise mass and redshift estimates.

%%%%%%%%%%%%%%%%%%%%%%%%%%%%%%%%%%%%%%%%%%%%%%%%%%%%%%%%%%%%%%%%%%%%%%%%%%%%%%%%%%%%%%%%%%
%%%%%%%%%%%%%%%%%%%%%%%%%%%%%%%%%%%%%%% Conclusions %%%%%%%%%%%%%%%%%%%%%%%%%%%%%%%%%%%%%%
%%%%%%%%%%%%%%%%%%%%%%%%%%%%%%%%%%%%%%%%%%%%%%%%%%%%%%%%%%%%%%%%%%%%%%%%%%%%%%%%%%%%%%%%%%

\section{Conclusions} \label{s:concl}
In this paper we have analysed the outputs of a set of six cosmological simulations with 
the aim of describing the effect of massive neutrinos on the LSS of the Universe, 
focusing on the SZ and X-ray properties of galaxy clusters. Our 
simulations follow the evolution of a very large comoving volume, 2\hmone\ Gpc per side, 
allowing us to describe accurately the halo mass function up to masses above $10^{15}$ 
\hmone\msun, in two different cosmological scenarios, one adopting parameters consistent 
with the latest \planck\ results \citep{planck14cp} and another with the \wmap\ results 
\citep{hinshaw13}, both considering three different values of the sum of neutrino masses: 
\smnut. The set of parameters used for these simulations is summarized in 
Table~\ref{t:cospar}. Starting from the outputs of each simulation we constructed a set 
of 50 light-cone catalogues of 10$^\circ$ per side and, by using known scaling relations, 
we computed the expected SZ and X-ray signals. This allowed us to test the effect of 
massive neutrinos on these observables, to compare them with \planck\ data and to 
investigate the capability of future cluster surveys of measuring \smnu\ and 
disentangling its degeneracy with other cosmological parameters, mainly \sigmae\ and 
\omegam. Our main results can be summarized as follows.

\begin{itemize}
\item[(i)] As expected, the effect of massive neutrinos is larger at higher masses 
and redshift. At $z=0$ neutrinos with masses \smnu=0.34 eV decrease the number of haloes 
by 20--40 per cent in the mass range \mfiveh$=10^{13.5-14.5}$\hmone\msun. While at $z=0$ 
this is almost equivalent to a change in cosmology from the \planck--\lcdmn\ model with 
\smnu$=0.34$ eV to a simple \wmap--\lcdm\ one, at higher redshift the effect of neutrinos 
is stronger.
\item[(ii)] The spatial correlation function of the haloes is also influenced by 
\smnu\ but with a smaller impact with respect to the mass function: about 15--20 per cent 
higher correlation for \smnu=0.34, with lower dependence on scale and redshift. 
Conversely, the effect associated with a change in \sigmae\ and \omegam\ is larger and 
scale dependent, allowing for a possibility of breaking the degeneracy of \smnu\ with 
these parameters.
\item[(iii)] The effect on the mass function translates into different expected 
counts of SZ detected clusters. Assuming a \planck--\lcdm\ cosmology \spt\ will be able 
to detect about 1100 clusters, reduced by 40 per cent when \smnu=0.34 eV. The same 
difference is found when computing the expected \planck\ detections in agreement with the 
findings of \cite{planck14cc}. However this is marginally in tension with the suggested 
value of \smnu$=(0.20\pm0.09)$ eV obtained when BAO constraints are included 
\citep{planck14cc}.
\item[(iv)] The global \ypar\ value is strongly influenced by the presence of massive 
neutrinos. We observe an approximate scaling of $(1-f_\nu)^{20}$ for the signal 
associated with galaxy clusters.
\item[(v)]  While the shape of the SZ power spectrum shows a weak dependence on \fnu, 
its normalization scales as $\left(1-f_\nu\right)^{25-30}$. The \spt\ and \act\ results 
at $\ell=3000$ are consistent with a \wmap--\lcdm\ cosmology. If \planck\ cosmology is 
assumed instead, a value of \smnu\ of about 0.34 eV is needed to fit with the 
data. We provide analytical fits to the power spectra (see equation~\ref{e:sknorm} and 
Table~\ref{t:psfit}).
\item[(vi)] Given the very high number of galaxy clusters above its flux detection 
limit, the \erosita\ full-sky survey has the potential to put constraints on \smnu, 
provided that cluster redshifts will be measured. However model uncertainties and the 
degeneracy of \smnu\ with both \sigmae\ and \omegam\ will make difficult to disentangle 
the two effects using its data alone.
\end{itemize}

Our work confirms how the LSS and galaxy clusters evolution are influenced by the value 
of the neutrino masses, the usefulness of cosmological analyses in measuring \smnu\ and 
the necessity of extending the traditional \lcdm\ concordance model to a more complete 
and accurate \lcdmn\ cosmology. With the increasing number of upcoming surveys, 
statistical uncertainties will not constitute the main limitation in providing 
strong limits on the neutrino masses using galaxy clusters. On the other hand, 
degeneracies with other cosmological parameters and with the modelling (mainly 
scaling laws and redshift evolution) will require the combination of both CMB and LSS 
data coming from different observable quantities, allowing us to probe different epochs 
of structure formation. A list of possible new types of analyses that may provide 
constraints on neutrino mass includes redshift-space distortions, CMB-weak lensing 
correlation, and kinetic SZ power spectrum. We plan to investigate these further neutrino 
probes and their cross-correlations in the near future by exploiting the
``Dark Energy and Massive Neutrino Universe'' (DEMNUni) simulation set (Carbone et al., 
in preparation, Castorina et al., in preparation), characterized by the same volume, but 
with a particle mass resolution about one order of magnitude larger than for the 
simulations adopted in the present work.

\section*{Acknowledgements}
Most of the computations necessary for this work have been performed thanks to the 
Italian SuperComputing Resource Allocation (ISCRA) of the Consorzio Interuniversitario 
del Nord Est per il Calcolo Automatico (CINECA). MR and LM acknowledge financial 
contributions from contract ASI/INAF I/023/12/0, from PRIN MIUR 2010-2011 `The dark 
Universe and the cosmic evolution of baryons: from current surveys to Euclid' and from 
PRIN INAF 2012 `The Universe in the box: multiscale simulations of cosmic structure'. CC 
acknowledges financial support to the `INAF Fellowships Programme 2010' and to the 
European Research Council through the Darklight Advanced Research Grant (\#291521). We 
thank an anonymous referee that contributed to improve the precision of our results and 
the presentation of our work. We thank M.~Brusa, S.~Ettori, F.~Marulli, F.~Finelli, 
F.~Petracca and M.~Sereno for useful suggestions and discussions. We are particularly 
grateful to M.~Viel for allowing us to access the non-public version of \gadgetthree\ 
modified to account for massive neutrino particles. We warmly thank A.~Saro and J.~Liu 
for providing us the data on the \spt\ noise.

\bibliographystyle{mn2e}
\newcommand{\noopsort}[1]{}

\label{lastpage}
\end{document}